\documentclass[preprint,prd]{revtex4}%

\usepackage{amsmath, amssymb, amsfonts, amsthm}
\usepackage{epsfig}
\usepackage{graphicx}  % standard LaTeX graphics tool
\usepackage{color}     % for color figures
\usepackage{lscape}
\usepackage{color}
\begin{document}

\title{ Modified Chaplygin Gas Cosmology from Geometrothermodynamics}

\author{Hachemi B. Benaoum$^1$ and Hernando Quevedo$^{2,3,4}$} 
\email{hbenaoum@sharjah.ac.ae,quevedo@nucleares.unam.mx}
\affiliation{
$^1$ Department of Applied Physics and Astronomy, 
Univesity of Sharjah, United Arab Emirates \\
$^2$Instituto de Ciencias Nucleares, Universidad Nacional Aut\'onoma de M\'exico,
 AP 70543, Ciudad de M\'exico 04510, Mexico\\
$^3$Dipartimento di Fisica and ICRANet, Universit\`a di Roma ``La Sapienza",  I-00185 Roma, Italy\\
$^4$Institute of Experimental and Theoretical Physics, 
	Al-Farabi Kazakh National University, Almaty 050040, Kazakhstan}

\begin{abstract}
The modified Chaplygin gas (MCG) cosmological model is derived by using the geometrothermodynamics (GTD)
formalism. We show that the MCG corresponds to a system with internal thermodynamic interaction and describes the
current accelerated expansion of the universe. Moreover, we also show that the MCG can be interpreted 
as a unified model for dark energy
and dark matter that is perfectly consistent with the current SNe observations and CMB anisotropy measurements.

\end{abstract}

\maketitle

%{\bf PACS numbers}: 98.80.-k, 98.80.Cq,98.80.Jk, 95.36.+x
%\begin{minipage}[h]{14.0cm}
%\end{minipage}
%\vskip 0.3cm \hrule \vskip 0.5cm
%%%%%%%%%%%%%%%%%%%%%%%%%%%%%%%%%%

%\newpage
%%%%%%%%%%%%%%%%%%%%%%%%%%%%%%%%%%%%%%%%%%%%
%%%%%%%%%%%%%%%%%%%%%%%%%%%%%%%%%%%%%%%%%%%%
\section{Introduction} 
\label{sec:int}

A growing number of observational data indicate that the observable universe is undergoing a phase of accelerated
expansion~\cite{per1}-\cite{spergel}. The most popular explanation for these unexpected observations states that 
the source of this cosmic acceleration is due to an unknown dark energy component
with a negative pressure, which dominates the universe at recent cosmological time.
Several mechanisms have been proposed to describe the physical nature of this dark energy component; among
them, the single-component fluid known as Chaplygin gas has attracted a lot of interest in recent times~\cite{peebles}-\cite{deffayet}.
Many variants of the Chaplygin gas model have been proposed in the literature. A further general model named
modified Chaplygin gas (MCG) has been introduced by Benaoum~\cite{ben1} and obeys the following equation of state: 
\begin{eqnarray}
p & = & A \rho - \frac{B}{\rho^{\alpha}} 
\end{eqnarray}
where $A, B$ and $\alpha$ are universal positive constants. \\

The equation of state with $A = 0$ and $\alpha = 1$ known as Chaplygin gas (CG) was first introduced by Chaplygin to
study the lifting forces on a plane wing in aerodynamics. Its generalized form with $A = 0$ and $\alpha > 0$ is known as
the generalized Chaplygin gas (GCG) was first introduced by Kamenshchik et al.~\cite{kame2} and Bento et al.~\cite{bento}. \\

The MCG has the remarkable property of describing the dark sector of the universe (i.e. dark energy and
dark matter) as a single component that acts as both dark energy and dark matter. It interpolates from a matter-
dominated era to a cosmological constant-dominated era. It is well known that the CG, MCG and theirs variants
have been extensively studied in the literature~\cite{deb}-\cite{ben3}. Several papers discussing various aspects of the behavior
of MCG to reconcile the standard cosmological model with observations have been considered~\cite{xu}-\cite{fabris}. \\

Geometrothermodynamics (GTD) is a formalism that has been developed during the past few years to describe
ordinary thermodynamics by using differential geometry~\cite{quevedo}. This formalism has found diverse applications in 
black hole thermodynamics \cite{bht}, relativistic cosmology \cite{cosm}, mathematical chemistry \cite{chem}, and others.

In this work we derive the MCG cosmological model
by means of GTD. In particular, we combine thermodynamic and geometric considerations in the GTD
framework, and find out the equation of state of the MCG which can explain the recent cosmological observations.
This work is organized as follows. In Section 2, we briefly review the fundamentals of GTD. Then, in Section 3,
we present the MCG cosmological model that follows from a GTD system with thermodynamic interaction. In
Section 4, we study the MCG cosmological model by using Type Ia supernovae observational data from Union 2.1
compilation. Finally, in Section 5, we discuss our results.

%%%%%%%%%%%%%%%%%%%%%%%%%%%%%%%%%%%%%%%%%%%%%%%%%%%%%%%%%%%%%%%%%% 
\section{Geometrothermodynamics}
\label{sec:gtd}

Geometrothermodynamics  is a differential geometric formalism for thermodynamics. The starting point of this formalism is 
a $(2 n + 1)-$dimensional contact manifold 
${\cal T}$, called phase space, which is endowed with a Riemanian metric $G$ and a contact $1$-form $\Theta$.
A set of coordinates 
$\{Z^A \}_{A=1, \ldots, 2n+1} = \{\Phi, E^a, I^a \}_{a=1, \ldots,n}$ are introduced where $\Phi$ represents the thermodynamic potential 
and $E^a$ and $I^a$  represent extensive and intensive thermodynamic variables, respectively.

An important property of GTD is that all its geometric objects are constructed such that they are invariant with respect to 
Legendre transformations. In classical thermodynamics, this is equivalent to saying that the physical properties of a given 
thermodynamic system do not depend on the choice of thermodynamic potential used for its description. A particular 
GTD metric $G$ which is invariant with respect to partial and total Legendre transformations can be written as (summation over all repeated
indices is implied)
\begin{eqnarray}
G & = & \Theta^2 + \Lambda \left(E_a I_a  dE^a dI^a \right) 
\end{eqnarray}
where the fundamental Gibbs $1$-form is $\Theta = d \Phi - \delta_{ab} I^a dE^b$ and $\Lambda$ is a real constant, which can be set equal to one without loss of generality.

The equilibrium $n$-dimensional submanifold ${\cal E} \in {\cal T}$ is defined by the smooth map, 
\begin{eqnarray} 
\varphi : & {\cal E} & \longrightarrow {\cal T} \nonumber \\
       & \{E^a \} & \longrightarrow \{\Phi (E^a), E^a, I^a (E^a) \}
\end{eqnarray}
such that the condition $\varphi^{\star} (\Theta) = 0$ is satisfied, implying that the first law of thermodynamics is 
satisfied on ${\cal E}$. 
Applying the pullback $\varphi^{\star}$ to the metric $G$, we get the induced thermodynamic metric $g$ given by: 
\begin{eqnarray}
g = \varphi^{\star} (G) &=& \left( E_a \frac{\partial \Phi}{\partial E^a} \right) 
~\frac{\partial^2 \Phi}{\partial E^b \partial E^c} \delta^{ab} dE^a dE^c 
\end{eqnarray}

According to the GTD prescription, one only needs to specify the fundamental equation $\Phi = \Phi (E^a)$ in order to find explicitly the metric $g$ of the equilibrium submanifold ${\cal E}$.  
% Notice that the summation inside the conformal factor must be performed in such a way that the resulting sum  leads to the thermodynamic potential $\Phi$.

\section{Modified Chaplygin Gas from GTD}

We choose the entropy $S = S \left(U,V \right)$ to be the thermodynamic potential and $U$, $V$ to be the extensive variables. Then, the corresponding 
thermodynamic phase space ${\cal T}$ is a five dimensional space endowed with the set of independent coordinates 
$Z^A = \{S, U, V, \frac{1}{T}, \frac{P}{T} \}$. 

The Gibbs' fundamental $1$-form $\Theta_S$ is given by: 
\begin{eqnarray}
\Theta_S & = & d S - \frac{1}{T} d U - \frac{P}{T} d V ~~~.
\end{eqnarray} 
By defining the space of equilibrium states ${\cal E}$ by $\varphi^{\star}_S (\Theta_S ) = 0$, we obtain both the first law of thermodynamics, 
\begin{eqnarray}
d S = \frac{1}{T} d U + \frac{P}{T} d V \ ,
\end{eqnarray}
and the equilibrium conditions, 
\begin{eqnarray} 
\frac{\partial S}{\partial U} & = & \frac{1}{T} \nonumber \\
\frac{\partial S}{\partial V} & = & \frac{P}{T} ~~~~.
\label{eqc}
\end{eqnarray}

To construct the modified Chaplygin gas (MCG) model by means of GTD, we consider the fundamental equation
\begin{eqnarray}
S & = & c_0 \ln V + \frac{c_1}{1+\beta} ~\ln \left( U^{\alpha +1} + c_2 V^{\beta +1} \right)\ ,
\label{feq}
\end{eqnarray}
where $c_0$, $c_1$, $c_2$, $\alpha $ and $\beta $ are real constants. This function is  a solution of the Nambu-Goto system of
differential equations
\begin{equation}
 \Box Z^A = \frac{1}{\sqrt{{\rm det}(g)}}
\left( \sqrt{{\rm det}(g)} g^{ab} Z^A_{,a} \right)_{,b} + \Gamma^{A}_{\, BC} Z^B_{,b} Z^C_{,c} g^{bc}=0
\end{equation}
where $\Box$ is the d'Alembert operator and $\Gamma^{A}_{\, BC}$ are the Christoffel symbols associated with the 
metric  $G_{AB}$ of the phase space. The above system of differential equations follow from the variational principle
$ \delta \int_{\cal E} \sqrt{{\rm det}(g)} d^n E =0$, which implies that the equilibrium space ${\cal E}$ with metric $g$ 
constitutes an extremal subspace of the phase space ${\cal T}$.

According to Eq.(3), the induced metric in the space of equilibrium states ${\cal E}$ is given as follows
\begin{eqnarray}
g & = & g_{U U} dU^2 + 2 g_{U V} dU dV + g_{V V} dV^2 \ ,
\end{eqnarray}
where the components of the thermodynamic metric $g$ can be expressed as
\begin{eqnarray} 
g_{U U} & = &  c_1^2 (1+ \alpha )^2 ~\frac{\alpha  c_2 V^{1+ \beta } - U^{1 + \alpha }}{(1+ \beta)^2 \left(U^{1+\alpha } + c_2 V^{1+ \beta } \right)^2} \ ,\nonumber \\
g_{V V} & = &  - \frac{1}{V^2} \left(c_0 + c_1 c_2 \frac{V^{1+\beta }}{U^{1+\alpha } + c_2 V^{1+ \beta }} \right) 
\left(c_0 + c_1 c_2 V^{1+\beta } \frac{c_2 V^{1+\beta } - \beta  U^{1+\alpha }}{\left(U^{1+\alpha } + c_2 V^{1+\beta } \right)^2} \right)\ , \nonumber \\
g_{U V} & = & - \frac{ (1+ \alpha ) c_1 c_2 U^{\alpha } V^{\beta }}{2 (1+\beta) \left(U^{1+\alpha } + c_2 V^{1+ \beta } \right)^2} \left(c_0 + 
\frac{c_1 (1+\alpha ) U^{1+\alpha } + (1+\beta ) c_2 V^{1+\beta }}{U^{1+\alpha } + c_2 V^{1+\beta }} \right) \ .
\end{eqnarray} 

Using the induced metric, we obtain the scalar curvature for the particular case $\beta= \alpha$ as: 
\begin{eqnarray}
R & = & \frac{N \left(U,V \right)}{D \left(U,V \right)} ~~~~~~,
\end{eqnarray} 
where
\begin{eqnarray}
N \left(U,V \right) & = & 
-8 (\alpha +1)^2 c_2 V^{\alpha +1} \left(c_2 V^{\alpha +1}+U^{\alpha +1}\right)^3 (c_2^3 (c_0+c_1) \nonumber \\
& &\times \left[(3 \alpha -5) c_0^2+(9 \alpha -5) c_0 c_1+4 \alpha  c_1^2\right] U^{\alpha +1} V^{3 (\alpha +1)} %\right. 
\nonumber \\
& & %\left. 
+ c_0 c_2 \left[(\alpha -7) c_0^2-(\alpha -1) c_1^2-2 c_0 c_1 \right] U^{3 (\alpha +1)} V^{\alpha +1}  %\right. 
\nonumber \\
& & %\left. 
-2 c_0^2 (c_0-c_1) U^{4 (\alpha +1)}+c_2^4 (c_0+c_1)^2 \left[(\alpha -1) c_0+4 \alpha  c_1 \right] V^{4 (\alpha +1)} \nonumber \\
& & +3 c_0 c_2^2 (c_0+c_1) \left[(\alpha -3) c_0+(\alpha -1) c_1 \right] U^{2 (\alpha +1)} V^{2 (\alpha +1)} ) \nonumber \\
D \left(U,V \right) & = &  c_1  (c_2^2 \left[ c_0^2(\alpha ^2+14 \alpha -11) -2 c_0 c_1 (\alpha ^2-6 \alpha +5) + c_1^2 (\alpha ^2+6 \alpha +1) \right] ~U^{2 (\alpha +1)} V^{2 (\alpha +1)}
%\right. 
\nonumber \\
   &  &%\left. 
	+ c_2^4 (c_0+c_1)^2 (\alpha ^2+6 \alpha +1) ~V^{4 ( \alpha +1)} \nonumber \\
   &  &   +  2 c_2^3 (c_0+c_1) \left[ c_0 (\alpha ^2+8 \alpha -1)- c_1 (\alpha -1)^2 \right] U^{\alpha +1} ~V^{3 (\alpha +1)} 
	%\right. 
	\nonumber \\
   & & %\left. 
	-4 c_0^2 U^{4 (\alpha +1)}+4 c_0 c_2 \left[c_0 (\alpha -3) +c_1 (\alpha -1) \right] U^{3 (\alpha +1)} V^{\alpha +1} )^2  ~~~~~~~~~~.
\end{eqnarray}
The Legendre invariant scalar curvature is in general non-vanishing which indicates the presence of internal 
thermodynamic interaction. In Figures 1 and 2, we have shown the three-dimensional behavior of the thermodynamic scalar curvature $R$ as a function of the energy $U$ and the volume $V$ for different values of $\alpha$ and $\beta$. It is interesting to note that even for $\beta = \alpha =0$ the scalar curvature does not vanish because of the presence of the parameter $c_0$. \\ 

\begin{figure}[hbtp]
\centering
\epsfxsize=6cm
\centerline{\epsfbox{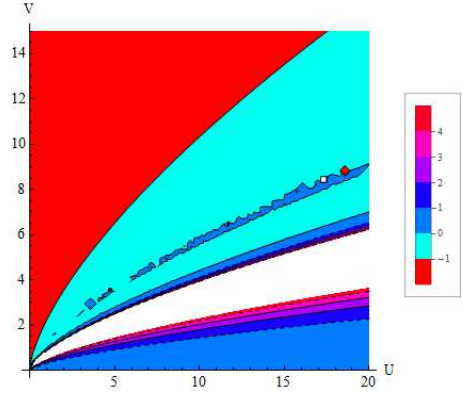}}
\caption{Contour plot of the curvature versus the energy $U$ and the volume $V$ for $\alpha = 1, \beta=2, c_0=0.3, c_1=0.9$ and $c_2=-0.5$.}
\end{figure}

 \begin{figure}[hbtp]
\centering
\epsfxsize=6cm
\centerline{\epsfbox{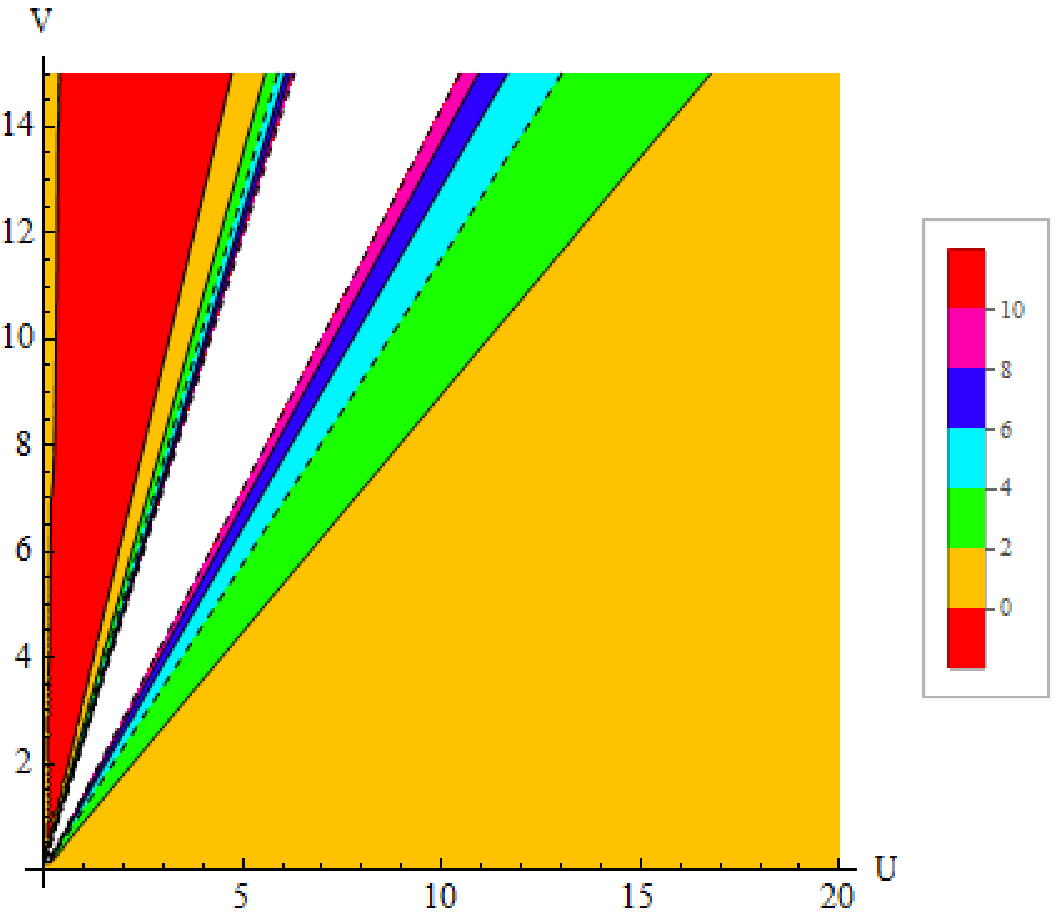}}
\caption{Contour plot of the curvature versus the energy $U$ and the volume $V$ for $\beta  = \alpha =0, c_0=0.3,c_1=1.$ and $c_2=-0.1$.}
\end{figure}

The conditions of equilibrium (\ref{eqc}) give
\begin{eqnarray}
\frac{\partial S}{\partial U} & = & \frac{c_1 (1+ \alpha )}{1+ \beta } \frac{U^\alpha }{U^{1+\alpha } + c_2 V^{1+\beta }} = \frac{1}{T} \label{con1} \\
\frac{\partial S}{\partial V} & = & \frac{c_0}{V} + c_1 c_2 \frac{V^\beta }{U^{1+\alpha } + c_2 V^{1+\beta }} = \frac{P}{T}\ ,
\end{eqnarray}  
which lead to an equation of state
\begin{eqnarray}
P & = & \frac{c_0}{c_1} \left( \frac{1+\beta }{1+\alpha } \right) \rho + c_2 \left( \frac{1+\beta }{1+\alpha } \right) \left(1+ \frac{c_0}{c_1} \right) \frac{V^{-(\alpha -\beta )}}{\rho^\alpha } .
\end{eqnarray} 

By defining $A = \frac{c_0}{c_1} \left(\frac{1+\beta }{1+\alpha }\right)$, $B = - c_2 \left(A + \frac{1+\beta }{1+\alpha } \right)$, the above equation can be written as
\begin{eqnarray}
P & = & A ~\rho - \frac{B V^{-(\alpha-\beta)}}{\rho^{\alpha}} 
\label{eos}
\end{eqnarray}
which for $\beta = \alpha$ is the MCG equation of state. From this equation one can see that for $\beta \neq \alpha$, it corresponds to the variable modified Chaplygin gas. 

One of the advantages of knowing the fundamental equation (\ref{feq}) explicitly is that it can be used to derive the complete thermodynamic information of the system. In particular, two thermodynamic quantities are important for the cosmological model under consideration, namely, the temperature, which follows from Eq.(\ref{con1}),
\begin{equation}
T= \frac{1}{c_1} \left( \frac{1+\alpha}{1+\beta} \right) \rho V  \left(1 + c_2 \rho ^{-\alpha-1} V^{\beta-\alpha}\right) ~~~~~,
\label{temp}
\end{equation}
and the heat capacity
\begin{equation}
C_V=-\frac{\left(\frac{\partial S}{\partial U}\right)^2}{\frac{\partial^2 S}{\partial U^2}} 
=  c_1\left( \frac{1+\alpha}{1+\beta}\right) \frac{1}{1-\alpha c_2 \rho ^{-\alpha-1} V^{\beta-\alpha}}\ .
\end{equation}
The arbitrary parameters which enter Eq.(\ref{temp})  must be chosen such that the temperature is always positive definite. Moreover, from  the heat capacity one can infer the phase transition structure of the system. Indeed, for a phase transition to take place 
($C_V\rightarrow \infty$), the condition 
\begin{equation}
\frac{c_2}{ \rho ^{\alpha+1} V^{\alpha-\beta}} = \frac{1}{\alpha}
\end{equation}
must be satisfied. If we also take into account that the temperature must be positive definite to be physically meaningful,
 then the above condition implies that
\begin{equation}
c_1\alpha (1+\beta) >0\ .
\end{equation}
This means that only for this particular choice of the free parameters a physical phase transition can occur. Since we have three different parameters available to satisfy this condition, one could in principle generate models with and without phase transitions. However, observations should impose additional limits on the range of values of the parameters entering each model.

\section{MCG Cosmology}
We consider a Friedmann-Lema\^itre-Robertson-Walker (FLRW) universe described by the following metric:
\begin{eqnarray}
ds^2 & = & - dt^2 + a^2 (t) \left[ \frac{dr^2}{1 - k r^2} +r^2 d \Omega^2_{D-1} \right] ~~~~~.
\end{eqnarray}
Here $a (t)$ is the scale factor of the universe and the curvature $k = 0, \pm 1$ describes spatially flat, closed or open spacetimes, respectively. \\

The Einstein field equations 
\begin{eqnarray}
R_{\mu \nu} - \frac{1}{2} g_{\mu \nu} R & = & T_{\mu \nu} 
\end{eqnarray}
for an one-component perfect fluid 
\begin{eqnarray}
T_{\mu \nu} & = & P ~ g_{\mu \nu} + \left( \rho + P \right)~ u_{\mu} u_{\nu} 
\end{eqnarray}
lead to the Friedmann equations which govern the evolution of the scale factor: 
\begin{eqnarray}
H^2 & = & \frac{2 ~\rho}{(D-1) (D-2)} - \frac{k}{a^2} \ ,
\end{eqnarray}
\begin{eqnarray}
\dot{H} & = & \frac{1}{D-2} ( \rho + P ) + \frac{k}{a^2}\ .
\end{eqnarray}
Moreover, the conservation law equation reads
\begin{eqnarray} 
\dot{\rho} + (D-1) ~H ( \rho + P) & = & 0 
\label{claw}
\end{eqnarray}   
with $H = \frac{\dot{a}}{a}$ being the Hubble parameter. Here, we assume that $c = 1$ and $8 \pi G = 1$. 

Assuming that $\rho$ as a function of time can be represented as a function of the scale factor $\rho(a)$, the conservation law can be integrated in a
straightforward manner. To this end, we assume the GTD equation of state (\ref{eos}) for the particular 
case $\beta=\alpha$. Then, the solution can be written as 
\begin{eqnarray}
\rho & = & \left[ \frac{B}{A+1} + \frac{C}{a^{(D-1) (\alpha+1) (A+1)}} \right]^{\frac{1}{\alpha+1}} ~~~,
\end{eqnarray}
where $C$ is a constant of integration. 

Consider now the particular case $D=4$. It is convenient to recast the above expression in the form
\begin{eqnarray}
\rho & = & \rho_0 \left[ A_s + (1- A_s) a^{-3 (A+1) (\alpha+1)} \right]^{\frac{1}{\alpha+1}} 
\end{eqnarray}
where $\rho_0$ is the present value of the dark energy density and $A_s = \frac{B \rho_0^{1-\alpha}}{1+A}$. \\ 
Now, using the redshift formula $z = \frac{1}{a} -1$, the Hubble parameter is obtained as: 
\begin{eqnarray}
\frac{H^2 \left(z \right)}{H^2_0} & = &  \left[\Omega_m (1+z)^3 + (1-\Omega_m) (A_s + (1 - A_s) (1+z)^{3 (A+1) (\alpha +1)})^{\frac{1}{\alpha+1}} \right] \ .
\end{eqnarray}
Here $\Omega_m = \frac{\rho_m}{H_0^2}$ is the present value of the dimension density parameter for matter and 
$H_0 \sim 72 km.s^{-1}. Mpc^{-1}$. 

The main evidence for the existence of dark energy was provided by Supernova Type Ia experiments. This means that the existence of dark energy is directly related to the redshift of the universe. Therefore, since 1995 two teams, the High Redshift Supernova Search and the Supernova Cosmology Project, have been working intensively and, as a result of their efforts, the have discovered several type Ia supernovae at high redshifts. The observations directly measure the distance modulus of a Supernovae and its redshift $z$. Here we will consider the recent observational data,including SNe Ia which consists of 557 data points and belongs to the Union 2.1 sample. From an observational point of view, the luminosity distance $d_L (z)$ defined as 
\begin{eqnarray}
d_L (z) & = & \left(1+ z \right) H_0 \int_0^z \frac{d z'}{H(z')} ~~~,
\end{eqnarray}
determines the dark energy density. Moreover, the distance modulus for Supernovae is given by:
\begin{eqnarray}
\mu (z) & = & 5 \log_{10} \left( \frac{d_L (z)/H_0}{1 Mpc} \right) + 25 ~~~. 
\end{eqnarray}
The best fit of the distance modulus $\mu(z)$ as a function of the redshift $z$ for the MCG cosmological model and the Supernova Type Ia Union 2.1 sample is shown in Fig.3. The best fit corresponds to $\Omega_m = 0.044, A=0.183, A_s=0.714$ and $\alpha=0.613$. From the curve, we see that the MCG model is in good agreement with the Union 2.1 sample data, indicating the validity of the model. 
\begin{figure}[hbtp]
\centering
\epsfxsize=6cm
\centerline{\epsfbox{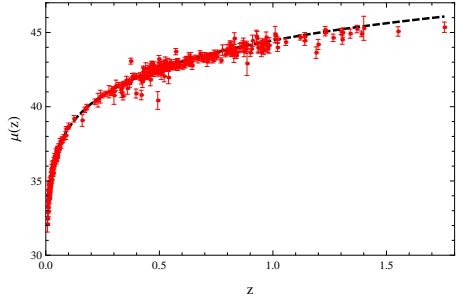}}
\caption{Plot of the distance modulus $\mu (z)$ versus the redshift $z$ .}
\end{figure}

\section{Conclusions}

In this work, we have used the formalism of GTD to construct a cosmological model for describing the dark sector of the universe. 
First, we
found a particular fundamental equation that determines an equilibrium space embedded in the
phase space by means of a map, satisfying the Nambu-Goto variational principle. It relates the entropy of a thermodynamic system with its internal energy and its volume. Moreover, it allows us to find all the physical properties of the system by using the standard laws and computational tools of classical thermodynamics.

The GTD
fundamental equation leads to an equation of state which is then used to integrate the Friedmann equations
of relativistic cosmology. By analyzing the physical properties of  the resulting model, 
we conclude that from GTD it is possible to obtain fundamental equations for thermodynamic systems 
that can be used to develop physically reasonable cosmological models. In particular, we applied the formalism of
GTD to construct a generalization of the MCG cosmological model. It turned out that this fluid corresponds to an equilibrium space with 
non-zero thermodynamic curvature, indicating
the presence of internal thermodynamic interaction. We performed a detailed analysis of the behavior of the
MCG and obtained that MCG, as a unified model for dark energy and dark matter, is perfectly consistent with
the current SNe observations.

%%%%%%%%%%%%%%%%%%%%%%%%%%%%%%%%%%%%%%%%%%%%%%%%%%%%%%%%%%%%%%%%%%%%%%%%%%%%%%%

\section*{Acknowledgements}

 This work was partially supported  by UNAM-DGAPA-PAPIIT, Grant No. 111617, and by the Ministry of Education and Science of RK, 
Grant No. BR05236322 and AP05133630.

\end{document}